# APPLICATION OF A SYSTEM OF INDICATORS FOR ASSESSING THE SOCIO-ECONOMIC SITUATION OF A SUBJECT BASED ON DIGITAL SHADOWS

**Olga G. Lebedinskaya**

Statistics department

Plekhanov Russian University of Economics

Moscow, Stremyanny pereulok 36

Lebedinskaya.OG@rea.ru

**ABSTRACT**

The development of Digital Economy sets its own requirements for the formation and development of so-called digital doubles and digital shadows of real objects (subjects/regions). An integral element of their development and application is a multi-level matrix of targets and resource constraints (time, financial, technological, production, etc.).

The volume of statistical information collected for a digital double must meet several criteria: be objective, characterize the real state of the managed object as accurately as possible, contain all the necessary information on all managed parameters, and at the same time avoid unnecessary and duplicate indicators ("information garbage").

The relevance of forming the profile of the "digital shadow of the region" in the context of multitasking and conflict of departmental and Federal statistics predetermined the goal of the work-to form a system of indicators of the socio-economic situation of regions based on the harmonization of information resources. In this study, an inventory of the composition of indicators of statistical forms for their relevance and relevance was carried out on the example of assessing the economic health of the subject and the level of provision of banking services.

***Keywords*** Big data, digital Economy, digital twin of the subject, economic health of the region, index of provision of banking services

## 1 Introduction

It is known that the digital shadow is able to predict the behavior of a real object only in the conditions in which the collection of big data was carried out, but it does not allow simulating situations in which the real object / product has not yet been exploited - the digital shadow has only a "memory property". Information and diagnostic digital twins provide monitoring and analysis of incidents, any abnormal behavior of the subject. That is, the algorithm operates: object - indicators - big data - predictive analytics. But in macroeconomic analysis, working with digital shadows and adapting statistical indicators for them are no less effective tools. An example of the digital shadow operation is indicative observation of the economic health of a region and an assessment of its provision with banking services.

## 2 Methodology

The authors ' team conducted an inventory of the composition of indicators of forms for their relevance and demand, an assessment of the level of duplication of collecting indicators, the level of development of the electronic method of reporting on the example of assessing the economic health of the region and the availability of banking services.

The set of indicators required for such analysis is contained in various information sources: specialized reference legal systems (databases), official Internet resources of public authorities (gov.ru) and information databases of state news agencies, in the collections Of the Institute for statistical research and Economics of knowledge of THE higher school of Economics.

The use of digital counterparts of subjects is almost limitless. Previously, the following macroeconomic indicators were used to assess the probability of a crisis in the economy: GDP, M2 aggregate, inflation, the share of loan debt in GDP, the share of Bank deposits in GDP, the spread in interest rates, etc. However, the problem of predictive significance of such studies is still open. Other studies are becoming relevant, which can only be carried out using the capabilities of digital counterparts of subjects. financial and credit market indicators should be primarily regional. The key parameter is the degree of adaptation of the banking sector to the region, although there is no doubt that the regional banking system is influenced by the external economic environment and the competitiveness of Russian enterprises ' products. Let's consider the possibilities of using digital twins of regions on the example of assessing the economic health of the region and their availability of banking services.

To develop indicators of the financial and credit market, you can use a system analysis technique based on quantitative indicators and qualitative interpretation of information. The basis for the formation of a General indicator of economic health based on six sub-indices (Sernyk, 2011):

1) Index of the physical volume of work performed by the type of activity "construction", including work performed by the household, in comparable prices;

2) Dynamics of investments in fixed capital by subjects of the Russian Federation

3) Dynamics of retail trade turnover per capita in the subjects of the Russian Federation;

4) Index of the physical volume of paid services to the population

5) the unemployment rate;

6) consumer and producer price Indices for the constituent entities of the Russian Federation

Other statistics were rejected as insignificant.

The analysis and assessment of the level of saturation of banking services in Russian regions was carried out using the adjusted methodology of the Bank of Russia also based on the use of sub-indices (Dyachenko, rodova, 2013). The reason for the need of such adjustments tied to its shortcomings: the relevance of the calculation is reduced if you consider only the credit institutions ' assets excluding the evaluation value of own funds (capital). In addition, it is also more appropriate to evaluate loans granted to individuals based on the population of the region rather than GRP per capita (Aminova, 2014), and to assess institutional security, on the contrary, using the GRP indicator of the region rather than the population.

Then the aggregate index of the region's banking services provision ($I_{rbsp}$) will be calculated using the following formula (1):

$$I_{rbsp} = \sqrt[8]{\mathrm{I}_1 \times \mathrm{I}_2 \times \mathrm{I}_3 \times \mathrm{I}_4 \times \mathrm{I}_5 \times I_6 \times \mathrm{I}_7 \times \mathrm{I}_8}$$

$I_1$ - Index of institutional provision of banking services;

$I_2$- Index of institutional provision of banking services;

$I_3$- Index of financial security of banking services by asset;

$I_4$- Index of financial security of banking services by capital;

$I_5, I$– accordingly, the Index of financial security of banking services provided to the population / legal entities to the GRP;

$I_7$- The index of development of savings business;

$I_8$ - Index of paid services.

All indicators are adjusted for the corresponding values for Russia.

## 3 Application

The study showed that in the country, despite all the efforts of the government, the economic health of the regions did not improve even during the period of sustained economic growth in some regions. The distribution of subjects according to the contribution of local indices to the formation of a composite indicator for the study period did not change significantly. So, just two indices - the index of growth of construction and growth in the volume of fixed capital investments in the sustainable leaders are remote subjects: Khanty-Mansi and Yamalo-Nenets Autonomous district, Sakhalin and Magadan region, Sakha Republic. The leading position of the Chukotka Autonomous Okrug in construction is associated with the implementation of state programs. It is noteworthy that the Federal cities of Moscow and St. Petersburg were not included in this group in absolute terms, but the dynamics of their changes is positive, and success in retail trade and providing paid services to the population is predetermined by their status.

Regional differentiation by the retail trade turnover index is more facilitated by the presence of a developed branch banking network in the subject than by the dynamics of inflation. Thus, the outsiders in this indicator are Kalmykia, Ingushetia, and the Karachay-Cherkess Republic. The spread of leaders and outsiders is shown in figure 1

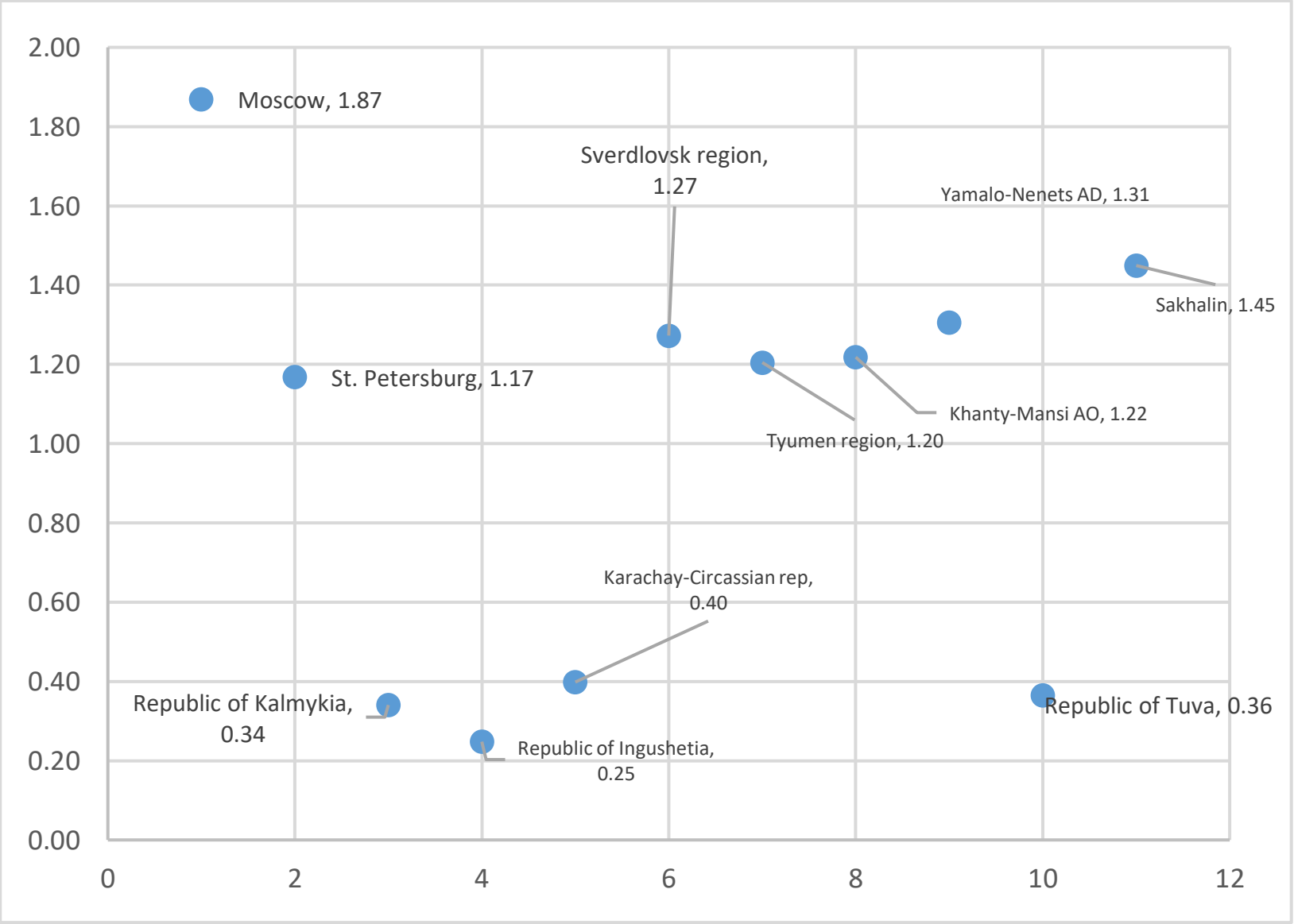

Figure 1-scatter Diagram of the retail trade turnover growth index values in the leading and outsider regions (compiled by the authors on the basis of data [5])

Two other local indices - the index of growth (decrease) in unemployment and the Index of growth (decrease) in the rate of inflation-also make a significant correction. The level of unemployment dynamics is quite contradictory. Some of the subjects with the maximum rate of unemployment growth according to the above-mentioned local indices look quite well. This is the Yaroslavl region (dynamics of the unemployment growth index – 129%, the Komi Republic-127%, the Samara region-123%). In these subjects, the deterioration of the economic situation is characterized by the influence of rather short-term factors. The other group of subjects (the Republic of Mari El, the Saratov region, the Kurgan region and the Republic of Buryatia, the Republic of Khakassia and Altai) have problems of a systemic nature, since all the previously considered local indices have minimal values, these are deeply depressed regions. For example, in 2019, the Republic of Mari El accumulated 16% of all state subsidies in the Volga-Kama region, the Saratov region-22.7% of the entire Volga-Ural region, and the Kurgan region – 61.6% of the Ural-Siberian region. In accordance with the adopted spatial development Strategy, all these subjects are included in the corresponding macro-regions as lagging behind in economic development, "satellites" of large agglomeration centers.

As for the dynamics of the local inflation growth index, the maximum values of the indicator were observed on the territory of raw materials and mining entities. The increase in this indicator is primarily due to inaccessibility and low population density. For this reason, but with a negative value, an increase in the indicator was recorded in almost all subjects of the Central Federal district.

Aggregation of data at the Federal district level significantly smooths the results obtained (Fig. 2).

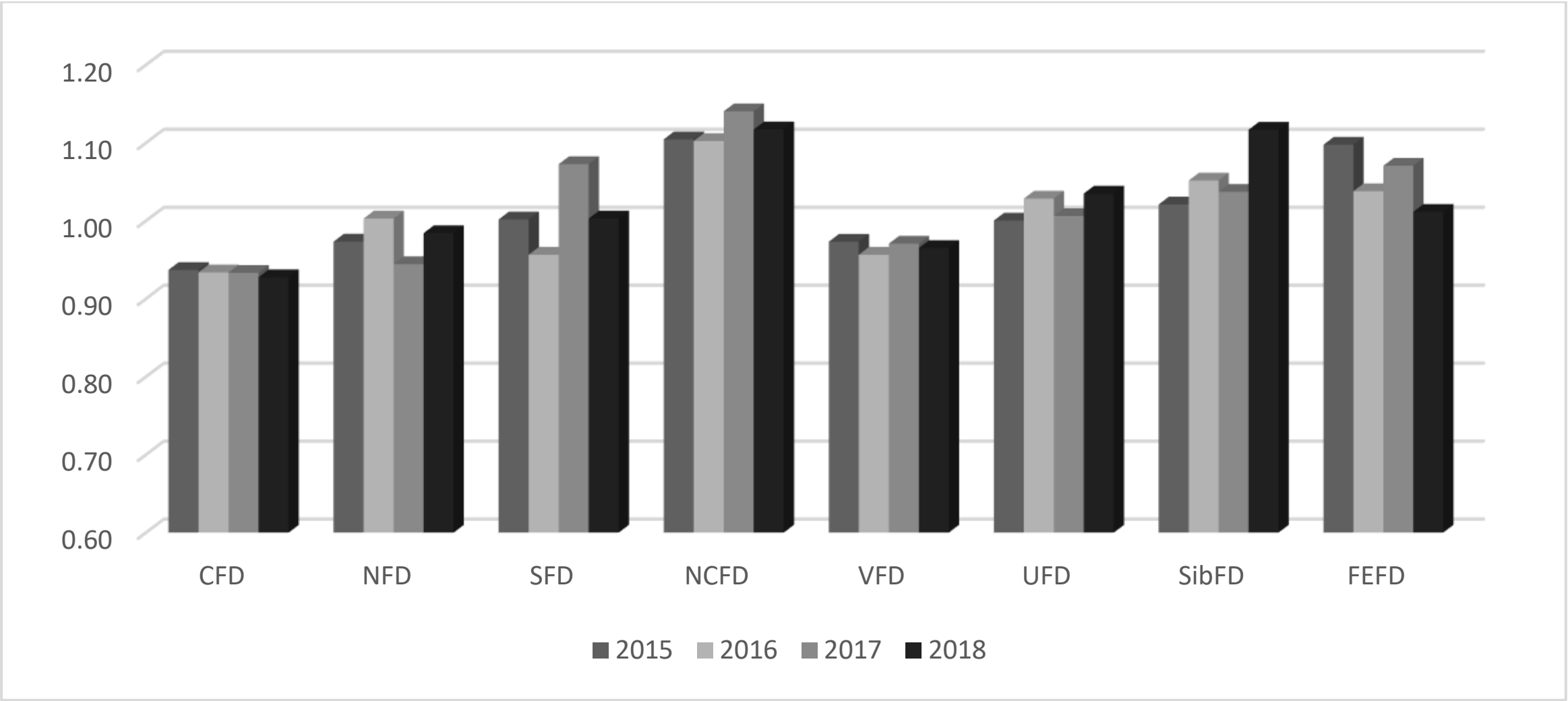

Figure 2-Dynamics of the indicator of economic health of Federal districts (compiled by the authors on the basis of data [5-7])

In the lagging republics of Tyva, Ingushetia, Kalmykia, Chechnya, as well as the TRANS-Baikal territory, Kurgan region. The last two subjects have positive dynamics, which may be due to a low comparison base. Moscow, Sakhalin and the Tyumen region are already traditional leaders.

When the research object changes from subjects to agglomerations, the assessment of economic health becomes more variable. The right-hand asymmetry increases (from 0.08 in 2015 to 0.12 in 2018), and the steepness of the distribution curve increases (from 2.21 to 3.18). The assessment of the dynamics of the completed grouping of subjects showed that there were no sharp "jumps" in the composition of the groups.

The economic health of subjects has a significant impact on its attractiveness for the credit market. For most Federal districts, the analysis revealed a close direct relationship with the index of banking services provision (R2=0.896). The wider the range of banking services provided in the region and the higher their volume, the more likely it is that high development indicators will be recorded in this territory. As in the assessment of economic health, data aggregation at the Federal district level significantly smooths the results.

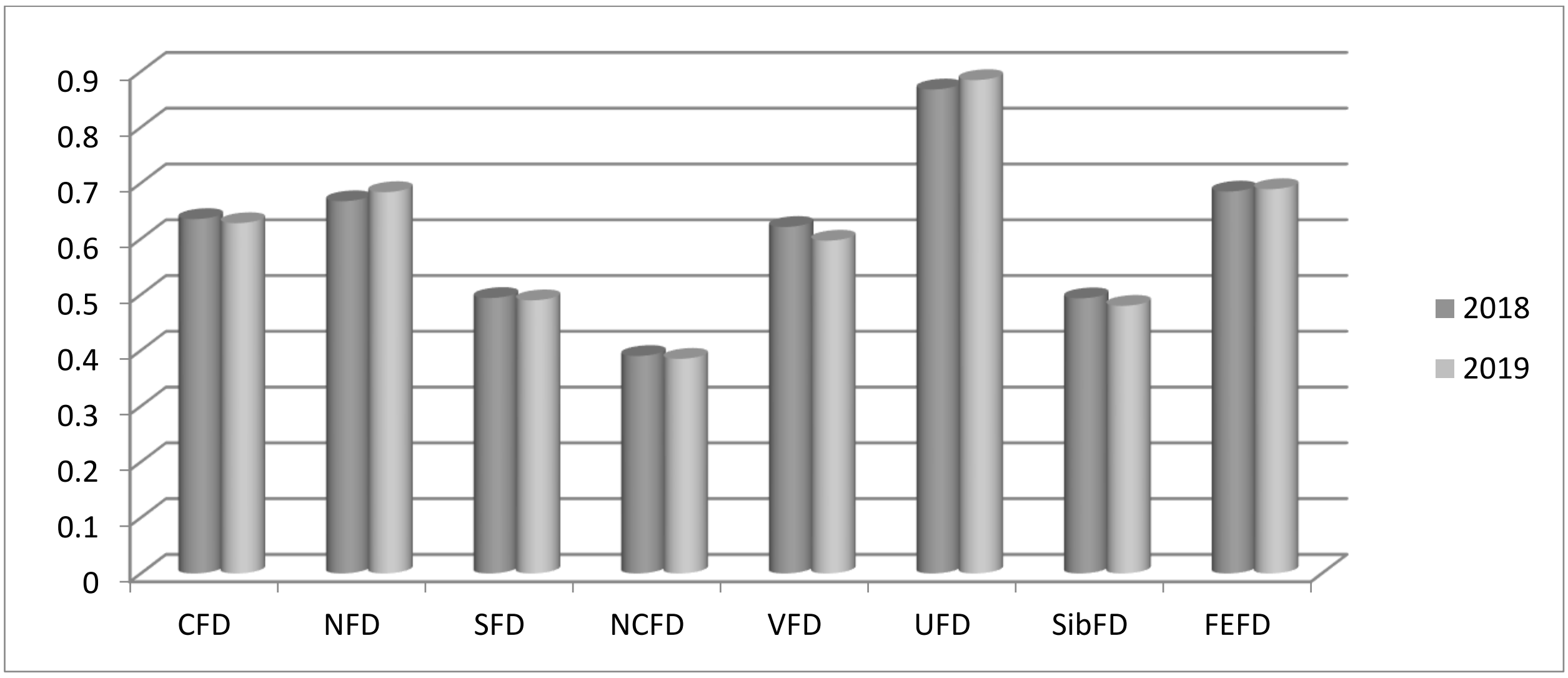

Figure 3-Dynamics of the index of banking services provision in Federal districts (compiled by the authors on the basis of data [5])

The influence of subindexes is uneven. This is confirmed by the scatter plot of index values, institutional banking services (by population) (Fig. 4) and the index of institutional banking services for the GRP (Fig. 5).

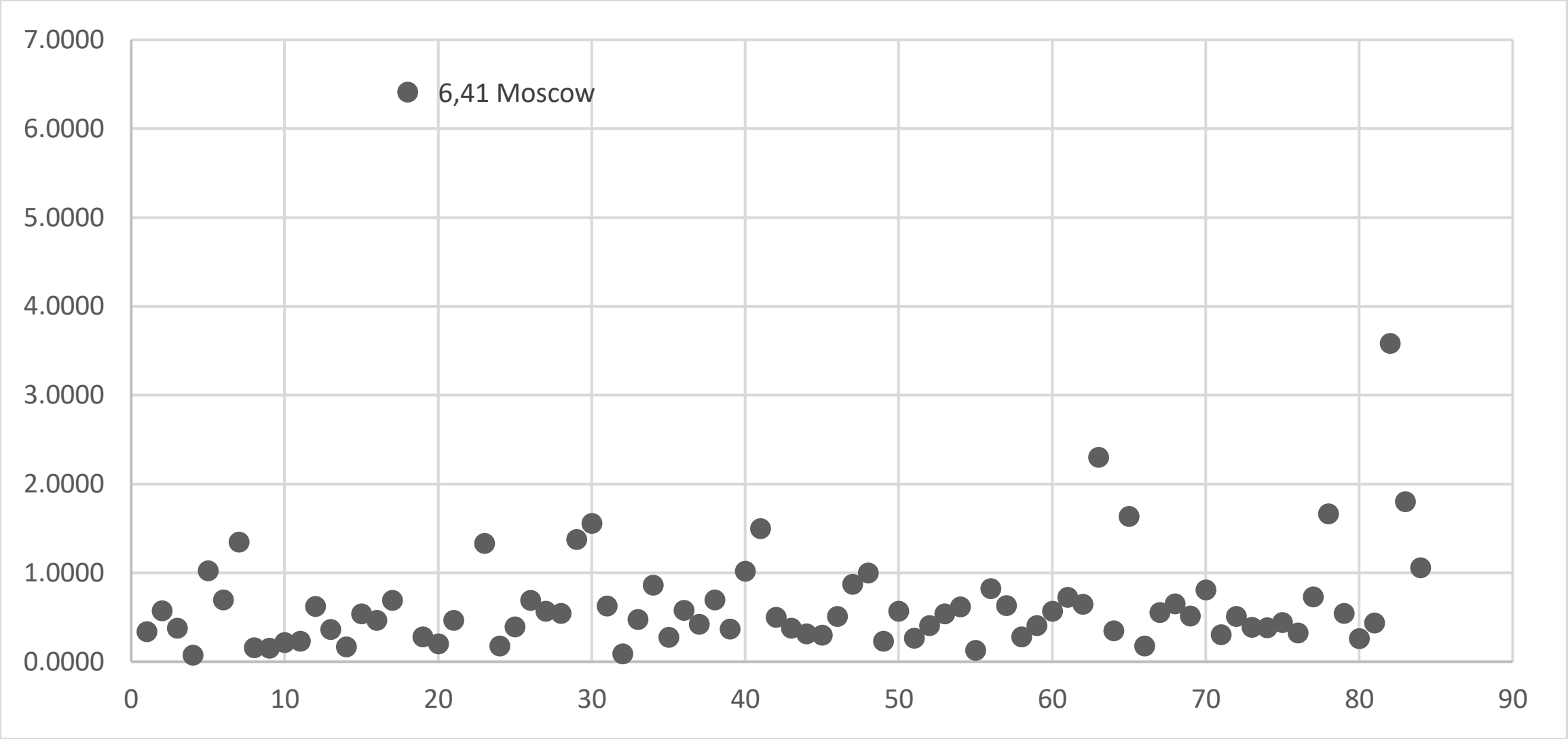

Figure 4-dispersion Diagram of the index of institutional provision of banking services (by population) (compiled by the authors on the basis of data [6])

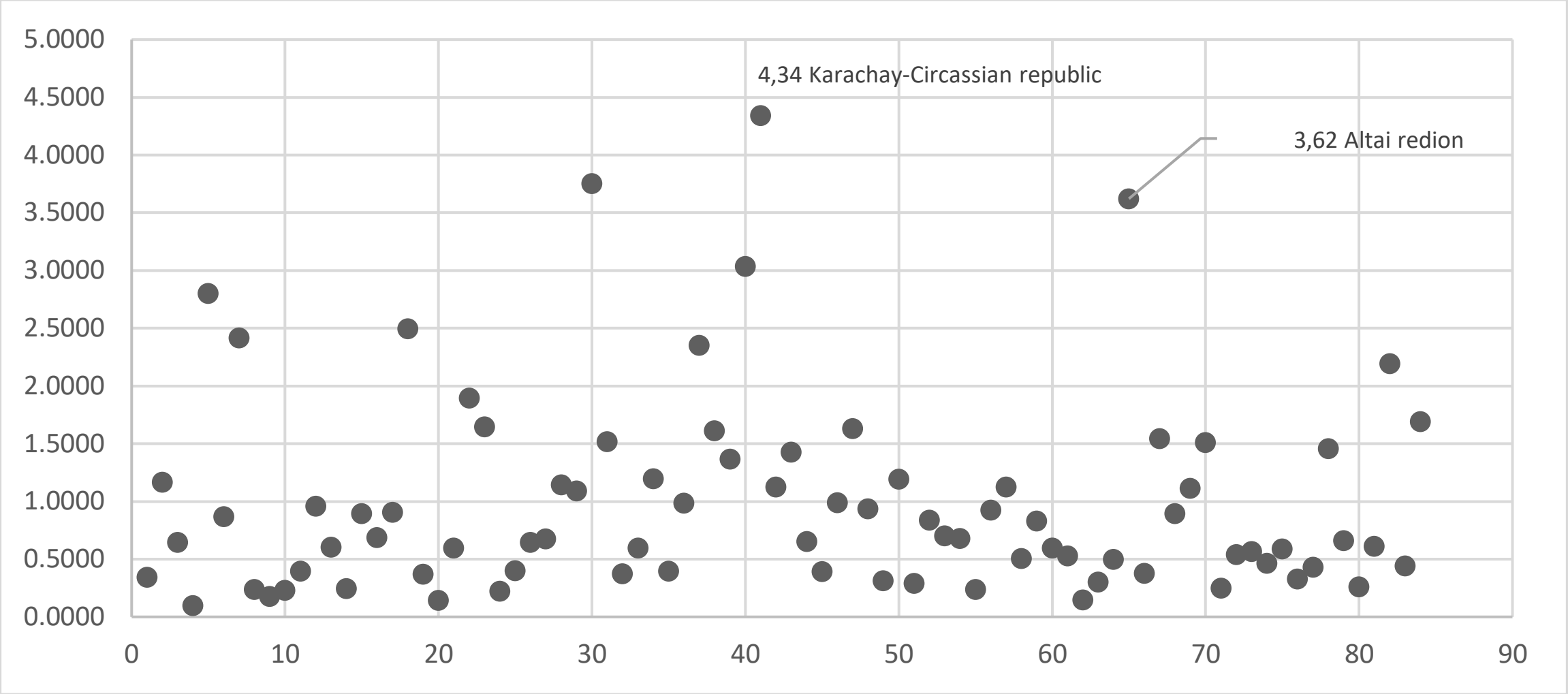

Figure 5-dispersion Diagram of the values of the index of institutional security in banking services (by GRP) (compiled by the authors based on data from

The group of leaders in the index of financial security of banking services in terms of loans to legal entities was formed by two types of entities with opposite motivation: a strategy for business development in regions with a high level of business activity (the largest agglomerations are Moscow and St. Petersburg) and a strategy for survival in depressed regions (the Karachay-Cherkess Republic and the Republic of Mari El).

High values of the index of financial security of banking services in terms of loans to individuals in remote regions (Nenets Autonomous Okrug, Khanty-Mansi, Yamalo-Nenets AO, Sakha Republic, Magadan oblast and Sakhalin) explain two reasons for the linkages between the developed commodity sector, ensuring increased income levels and cost of living and policy of banks to facilitate the citizens access to credit.

This statement can be confirmed by data from other information bases, which once again confirms the need to form digital doubles by aggregating information from different sources. So, according to the portal banki.ru in terms of the absolute volume of loans per economically active citizen, the Yamalo-Nenets Autonomous district is the leader with a debt volume of 478 thousand rubles. The second, third and fourth places are occupied by the Khanty-Mansi Autonomous Okrug, Yakutia and the Nenets Autonomous Okrug, which have a debt level per person of 445, 389 and 359 thousand rubles, respectively.

The highest level of development of the savings business is traditionally recorded in two Federal districts – Central and Volga. In almost all regions of the Central Federal district (with the exception of the Bryansk and Kursk regions), the savings development index values are higher than one: Voronezh region (1.22), Moscow region (1.15), Ryazan region (1.15), Tver region (1.17), Moscow (1.41)

According to the index of paid services (Fig. 5), the stable leaders are Crimea (5.15) and Sevastopol (2.86), the positive dynamics of which is explained by a sharp increase in the volume of services provided by hotels, accommodation facilities, travel companies, health centers, and cultural institutions. A significant contribution to the dynamics of the index and transport services was noted. Outsiders are also traditional. The lag of these subjects is primarily due to either low population density or difficult natural and climatic conditions. This group was formed by Sakhalin (0.45) and Chukotka Autonomous Okrug (0.56), Khanty-Mansi and Yamalo-Nenets Autonomous okrugs (0.25 and 0.17, respectively).

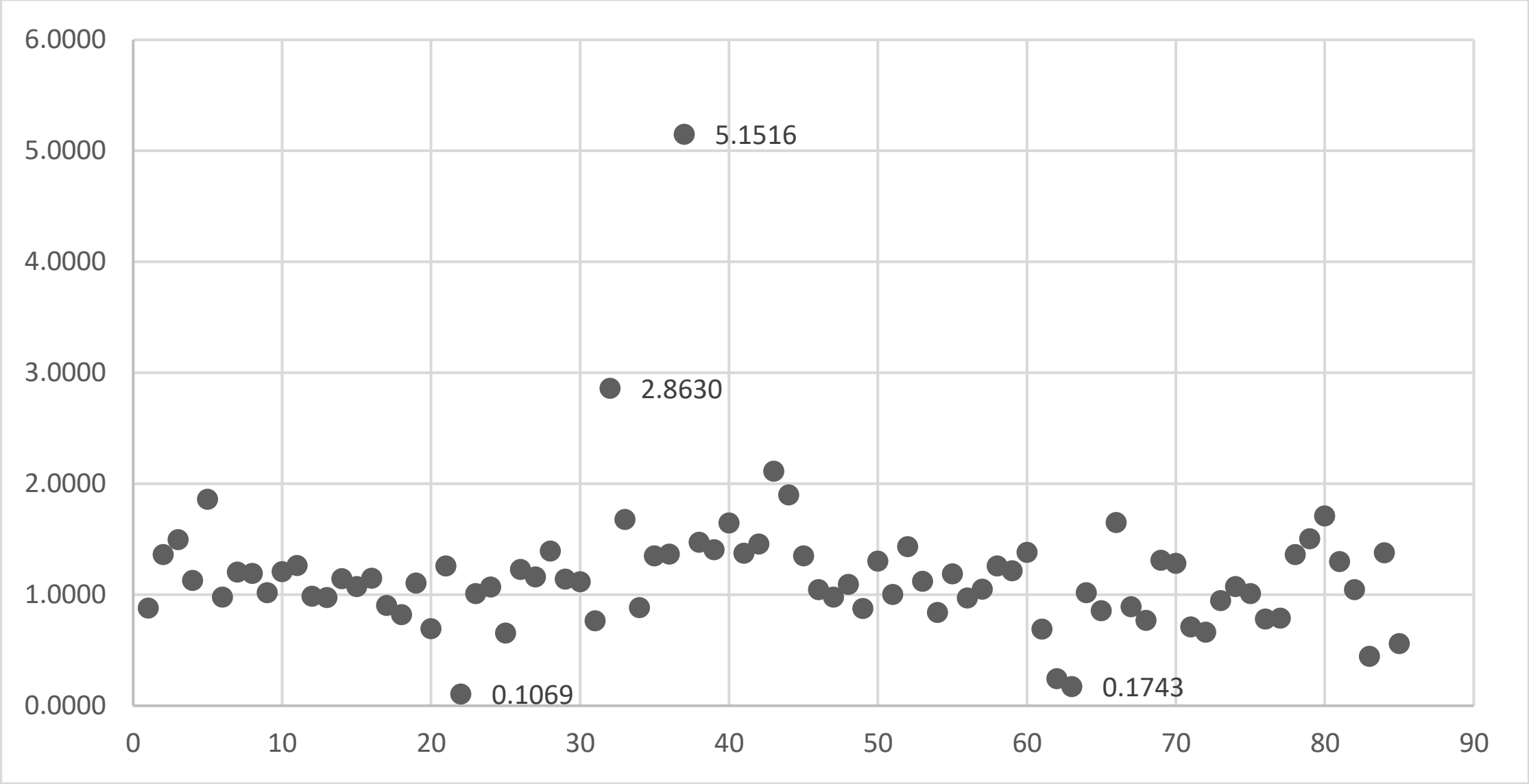

Figure 6-dispersion Diagram of the paid services index values (compiled by the authors based on data from [10])

A significant spread in the values of the banking services security index of subjects was revealed (coefficient of variation 0.763), and the asymmetry with the same slope is even more acute (E=9.33) (Fig. 6).

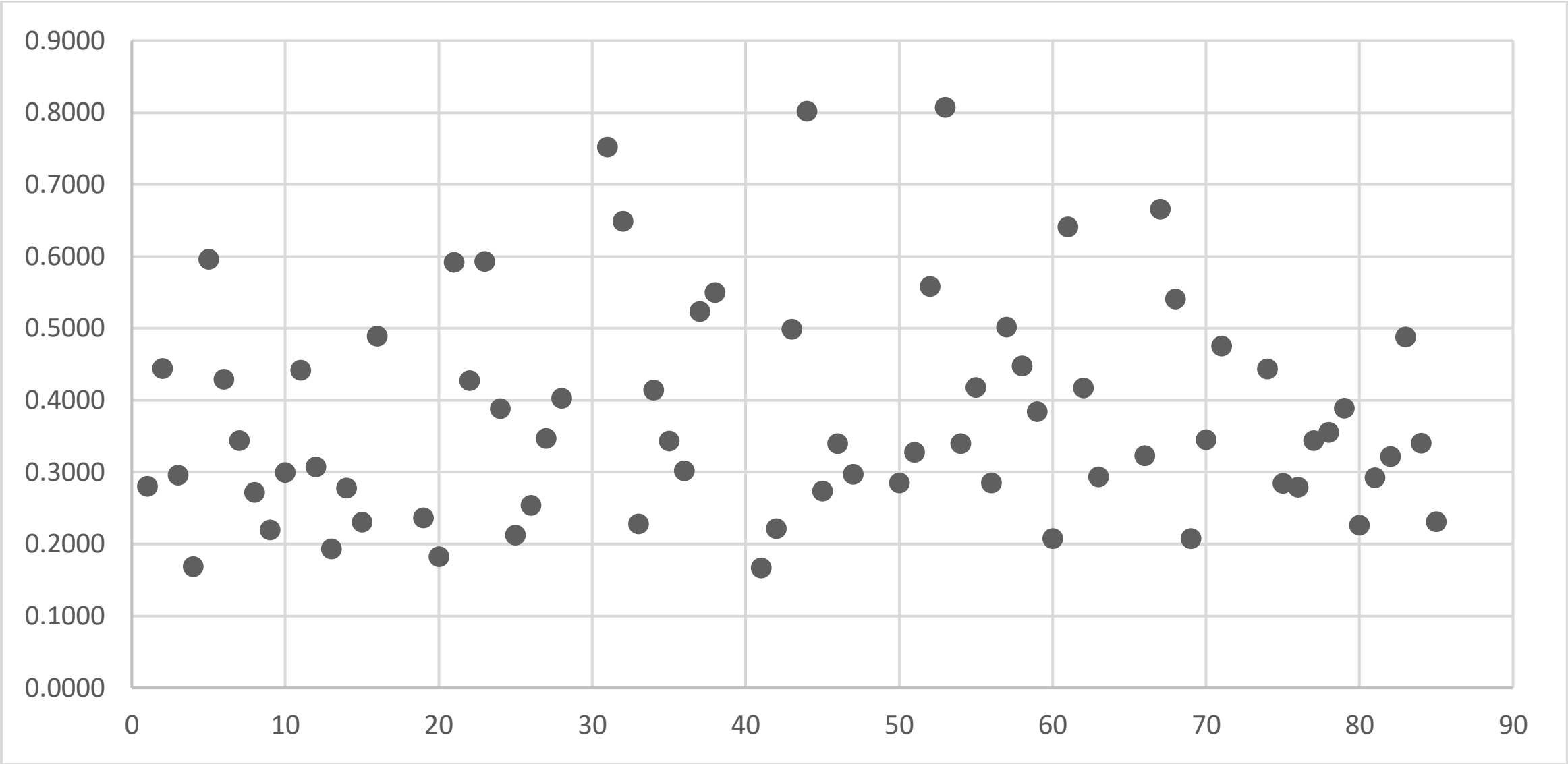

Figure 7-dispersion Diagram of the index of banking services provision of subjects in 2019.

In subjects included in the lower quartile, the index value does not exceed 0.19, in the upper quartile, the index values do not decrease less than 0.56. During the study period, the composition of outsiders did not change: Dagestan, Voronezh region, Komi Republic, Tambov region. The steady inclusion of the Voronezh region in this group, despite having the status of a key center of economic development, can be explained by the peculiarities of the agglomeration development of the territory and its proximity to Moscow.

However, in practice, there may be situations when the inverse relationship is fixed with high growth rates of the region's GRP (especially if it is ahead of the national average) and with a moderate increase in lending volumes, the value of the financial saturation indicator decreases. It is worth noting the relationship between the parameters of the credit, increase GRP and the time lag: from the moment of the loan enterprise to the development of these funds in full and its output at full capacity may take considerable time, after which the plant will have a direct impact on GRP.

The level of provision of banking services also determines the behavior of the subject in the financial sphere. The authors suggest that there are four types of regions. The first group is represented by regions with high incomes and a greater propensity of the population to save, these are "profit centers" with a high agglomeration potential (type I).

The second group includes regions with high rates of income differentiation, export-oriented specialization, and often insufficiently developed infrastructure

## 4 Discussion

The methods for assessing the provision of regions with banking services and assessing economic health presented in the article make it possible to evaluate in practice the effectiveness of working with the digital shadow of a region. The efficiency of working with digital shadows is increasing in the context of the transition to a new paradigm of territorial accounting (from federal districts to agglomeration development).

## Acknowledgements

This research was performed in the framework of the state task in the field of scientific activity of the Ministry of Science and Higher Education of the Russian Federation, project "Development of the methodology and a software platform for the construction of digital twins, intellectual analysis and forecast of complex economic systems", grant no. FSSW-2020-0008.